\documentclass[aps,prl,twocolumn,floats,showpacs]{revtex4}
\usepackage{epsfig}
\usepackage{graphicx}
\usepackage{amsmath}

\newcommand{\beq}{\begin{equation}}
\newcommand{\eeq}{\end{equation}}
\newcommand{\bea}{\begin{eqnarray}}
\newcommand{\eea}{\end{eqnarray}}
\newcommand{\eq}[1]{Eq. (\ref{#1})}

\newcommand {\ket}[1]{|\,{#1}\,\rangle}
\newcommand {\bra}[1]{\langle\,{#1}\,|}

\begin{document}

\title{Derivation of the relativistic ``proper-time'' quantum evolution
equations from Canonical Invariance}

\author{Moshe Shapiro}

\affiliation{Department of Chemistry, The University of British
Columbia, Vancouver, B.C. V6T 1Z3, Canada, and
Department of Chemical Physics, The Weizmann Institute,
Rehovot 76100, Israel.}


\begin{abstract}
Based on
1) the spectral resolution of the energy operator;
2) the linearity of correspondence between physical observables and quantum
self-adjoint operators; 3) the definition of conjugate coordinate-momentum
variables in classical mechanics; and 4) the fact that
the {\it physical} point in phase space remains unchanged under
(canonical) transformations between
one pair of conjugate variables to another, we are able to show that
$\bra{t_s}E_s\rangle$, the proper-time rest-energy transformation matrices,
are given as $a\exp(-iE_st_s/\hbar),$
from which we obtain the proper-time rest -energy evolution equation
$i\hbar{\partial\over \partial t_s}\ket{\Psi}=\hat{E_s}\ket{\Psi}$.
For special relativistic situations this equation can be reduced
to the usual $i\hbar{\partial\over \partial t}\ket{\Psi}=\hat{E}\ket{\Psi}$
dynamical equations, where $t$ is the ``reference time" and $E$ is the
total energy. Extension of these equations to {\it accelerating} frames is
then provided.
\end{abstract}


\maketitle

\section{Introduction}

Attempts at proving the dynamical equations of quantum mechanics (e.g.,
the time dependent Schr\"odinger equation), rather than
{\it assuming} them\cite{cohen}, as did Schr\"odinger himself\cite{schroed},
or {\it postulating}
the $H\rightarrow i\hbar
\partial/\partial t$ correspondence\cite{messiah,roman},
abound in the literature.
Most of these approaches
achieve this purpose by showing that the dynamical equations are equivalent
to other assumed
postulates, such as stochastic dynamics\cite{nelson,davies,pelce},
path-integrals\cite{dewitt,cheng,fiziev}
or Galilean symmetry\cite{fushchych}.
Some\cite{briggs} attempt to treat time within the framework of the
{\it time-independent} Schr\"odinger equation, as a semiclassical quantity
associated with the dynamics of
a large semiclassical bath coupled to the quantum system of interest.
This approach neglects the vast literature
and experimental evidence\cite{zewail} regarding the temporal
evolution of wave packets in {\it isolated} system which
cannot be described by the time-independent
Schr\"odinger equation alone.

Recently\cite{shapiro06} we have shown that
the $[\hat{q}_l,\hat{p}_l]=i\hbar$
commutation relations between any generalized coordinate operator $\hat{q}_l$
and its conjugate momentum operator $\hat{p}_l$
can be derived from the very definition of conjugate variables in
classical mechanics. The derivation made use of the invariance of the
free particles' kinetic
energy and other invariant operators
to (``canonical'') transformations between one pair of conjugate variables
and another. In the present paper we extend this approach to proving
the relativistic (and non-relativistic)
\beq
i\hbar{\partial\over \partial t_s}\ket{\Psi}=\hat{E}_s\ket{\Psi}
\label{energy}
\eeq
quantum evolution equations, with $t_s$ being the proper time
and $\hat{E}_s$ - the
rest-energy operator.

In the coordinate-momentum case the $\hat{p}=-i\hbar\partial/\partial x$
equation analogous to \eq{energy} is equivalent to the $[x,p]=i\hbar$
commutation relation. This cannot be easily done in the time-energy
case because of our seeming inability to construct a self-adjoint
operator for the ``time''.
In particular, it was claimed by Pauli\cite{pauli},
that if such an operator existed it would imply
that its conjugate operator, namely the Hamiltonian, would have a
purely continuous spectrum and be unbounded from below.
More specifically,
if there exists a self-adjoint time operator $\hat{t}$ conjugate to the
Hamiltonian $\hat{H}$ operator,
such that $[\hat{t},\hat{H}]=-i\hbar,$ then
if $\ket{\Phi_{E_i}}$ is
an eigenstate of $\hat{H}$ satisfying $\hat{H}\ket{\Phi_{E_i}}=
E_i\ket{\Phi_{E_i}}$,
$\ket{\Phi_{E_i-\beta}}\equiv\exp(i\beta\hat{t}/\hbar)\ket{\Phi_{E_i}}$
would also be an eigenstate
of  $\hat{H}$ with an eigenvalue $E_i-\beta$. Since $\beta$ is arbitrary
this would imply that the spectrum of $\hat{H}$ is continuous and
unbounded from below.

Subsequently it was shown\cite{galapon} that
Pauli's arguments were flawed and that for a bounded
time operator a conjugate Hamiltonian with a point spectrum {\it can} exist.
A simplified way of explaining this flaw is to say that
if $\langle x\ket{\Phi_{E_i}}$
is square-integrable, for an arbitrary $E_i-\beta,$
$\langle x\ket{\Phi_{E_i-\beta}}$ is not square integrable as it
diverges either as $x\rightarrow\infty$ or
as $x\rightarrow -\infty$. It is therefore not part of the spectrum
of the Hamiltonian.

In what follows we define a {\it bounded}, {\it self-adjoint} ``proper-time''
operator, based on
the classical ``proper-time''
in different inertial frames. This operator assumes the form
$\hat{t}_s\equiv t(1-\hat{v}^2_s/c^2)^{1\over 2}~,$
where $t$ is a {\it parameter} representing the time
measured in one of the frames, designated the ``reference'' frame,
and $\hat{v}_s$ is the velocity operator of
the $s$-frame.
We argue that in quantum mechanics there is a natural spread in the
observed $t_s$ values due to
the uncertainty in
the s-frame velocity.

After identifying below the variable conjugate to $t_s$
in relativistic mechanics as $-E_s$, where $E_s$ is the rest-energy of
particle $s$,
we make use of canonical invariance
to prove the
$\langle t_s|E_s \rangle=a\exp(-iE_st_s/\hbar)$
identity, a special case of which being
$\langle t|E\rangle=a\exp(-iEt/\hbar),$ where $E$ is the total energy,
leading to \eq{energy}.
Thus, together with the {\it ``ab-initio''}
derivation of the coordinate representation of
the momentum operator presented in Ref.\cite{shapiro06},
we now have a more firmly based, reduced number of
axioms, theory of relativistic
quantum evolution.

\section{Review of the proof of the
Cartesian $[\hat{x},\hat{p}]$ commutation
relations in relativistic quantum mechanics }
\label{xp}
In order to motivate what follows we briefly review part of
Ref.\cite{shapiro06} in which the
Cartesian $[\hat{x},\hat{p}]$ commutation
relations of relativistic quantum mechanics were derived.
We consider two free particles $A$ and $B$ of equal rest mass $m_A=m_B$
whose Cartesian extensions on the $x$ axis are denoted $x_A$ and $x_B$.
The momenta conjugate to these coordinates
are obtained from the free Lagrangian\cite{goldstein}
\beq
{\cal L}=-mc^2\left[(1-\beta^2_A)^{1\over 2}+(1-\beta^2_B)^{1\over 2}\right]
\label{rellag}
\eeq
where $\beta_{A(B)}=v_{A(B)}/c~,$ as
\beq
p_s=\partial{\cal L}/\partial v_s=mv_s/(1-\beta^2_s)^{1\over 2}~,s=A,B.
\label{ps}
\eeq

We now make a
canonical transformation to the
$X_{1(2)}=(x_A\pm x_B)/\sqrt{2}~,$
variables, with the associated velocities
$V_{1(2)}=\dot{X}_{1(2)}$,
being given as, $V_{1(2)}=(v_A\pm v_B)/\sqrt{2}~.$
Substituting the velocity relations in \eq{rellag} we have that
\beq
-{\cal L}/(mc)=
\left(c^2-{(V_1+V_2)^2\over 2}\right)^{1\over 2}
         +\left(c^2-{(V_1-V_2)^2\over 2}\right)^{1\over 2},
\label{tlagrange}\eeq
from which we obtain that
the momenta conjugate to $X_1$ and $X_2$ are given as,
\beq
P_1={\partial{\cal L}\over\partial V_1}
={p_A+p_B\over\sqrt{2}}~,~~
P_2={\partial{\cal L}\over\partial V_2}=
{p_A-p_B\over \sqrt{2}}~.
\eeq

We now consider the sum of
the {\it squares} of the energies of the two particles,
\beq
{\cal T}=T^2_A+T^2_B=
c^2\{2m^2c^2+p^2_A+p^2_B\}~.
\label{relt}
\eeq
It is easy to show
that ${\cal T}$ is invariant to the canonical
transformation to the $X_1$ and $X_2$
variables and can be written as  ${\cal T}=c^2\{2m^2c^2+P_1^2+P_2^2\}.$

Since $\hat{{\cal T}}$, the quantum operator corresponding to ${\cal T}$,
is a function of the
$\hat{p}_A$ and $\hat{p}_B$ operators, which commute between themselves,
and is also a function of the $\hat{P}_1$ and $\hat{P}_2$ operators,
which also commute between themselves,
it must commute with all four momenta,
\beq
[\hat{\cal T},\hat{p}_A]=[\hat{\cal T},\hat{p}_B]=
[\hat{\cal T},\hat{P}_1]=[\hat{\cal T},\hat{P}_2]=0~.
\nonumber\eeq
This fact, plus the separable form of $\hat{{\cal T}}$,
allows us to write the eigenstates of
$\hat{\cal T}$ in two different ways
\beq
\langle x_A|p_A\rangle\langle x_B|p_B\rangle=\langle X_1|P_1\rangle
\langle X_2|P_2\rangle~.
\label{1r}
\eeq

Choosing two particular momenta, $p_A=p$ and $p_B=-p$, and two
particular positions $x_A=x$ and $x_B=-x$, we have for these values that
$P_1=0~,P_2=\sqrt{2}p~,X_1=0~,X_2=\sqrt{2}x~.$
Equation (\ref{1r}) now assumes the special form,
\beq
\bra{x}p\rangle
\bra{-x}-p\rangle
=\bra{0}0\rangle
\bra{\sqrt{2}x}\sqrt{2}p\rangle~.
\label{3r}
\eeq
Obviously the $\bra{-x}-p\rangle$ amplitude is independent of the definition
of our coordinate system. Thus, if we re-define $-x$ to be $x$,
forcing by \eq{ps} (since the Lagrangian remains invariant to this
re-definition), $-p\rightarrow p,$ we obtain that
$\bra{-x}-p\rangle=\bra{x}p\rangle,$ and it follows from \eq{3r} that
\beq
\langle 0|0\rangle\langle\sqrt{2}x|\sqrt{2}p\rangle
=\left(\langle x|p\rangle\right)^2~,~{\rm or}~
\langle\sqrt{2}x|\sqrt{2}p\rangle'
=\left(\langle x|p\rangle'\right)^2
~,
\label{4r}
\eeq
where $\langle x|p\rangle'
\equiv \langle x|p\rangle/\langle 0|0\rangle.$

Defining
$\delta_p\equiv p/2^{n/2}~$,~ and $\delta_x\equiv x/2^{n/2}~$,
we have from \eq{4r} that
\beq
\langle x|p\rangle'=\langle 2^{n/2}\delta_x|2^{n/2}\delta_p\rangle'=
\left(\langle\delta_x|\delta_p\rangle'\right)^{2^n}.
\nonumber
\eeq
Hence,
\beq
\log \langle x|p\rangle'= 2^{n/2}\delta_x2^{n/2}\delta_p
\alpha'(\delta_x,\delta_p)= px \alpha'(\delta_x,\delta_p)
~,
\nonumber
\eeq
where
\beq
\alpha'(\delta_x,\delta_p)=
\log\langle\delta_x|\delta_p\rangle'/(\delta_p\delta_x)~.
\nonumber
\eeq
For every $p$ and $x$ values we can find an $n$ value, such that
$\delta_x$ and $\delta_p$ are sufficiently small so that
$\alpha'(,\delta_x\delta_p)$ is sufficiently close to its
limiting value
\beq
\alpha=
\lim_{\delta_p\rightarrow 0,\delta_x\rightarrow 0}
{\log\langle\delta_x|\delta_p\rangle'\over \delta_x\delta_p}
={\partial^2\log\langle x|p\rangle'\over \partial x\partial p}\Big|_{x,p=0}.
\nonumber
\eeq
Hence
\beq
\log \left(\langle x|p\rangle'\right)= \alpha px ~,{\rm or}~
\langle x|p\rangle= \langle 0|0\rangle\exp(\alpha px)
~.
\nonumber\eeq
In order for $\langle x|p\rangle$ to be normalizable
to $\delta(x-x')$, namely
\beq
\int\int dpdx' \langle x|p\rangle\langle p|x'\rangle=1~,
\label{delta}
\eeq
$\alpha$ must be a purely imaginary number
$\alpha=i\gamma$, (otherwise $\langle x|p\rangle$ would diverge, either as
$x\rightarrow\infty$ or as $x\rightarrow -\infty$).
By
identifying $\gamma$ with $1/\hbar$, we obtain that
\beq
\langle x|p\rangle=
\langle 0|0\rangle
\exp(i\gamma px) =
(2\pi\hbar)^{-1/2}
\exp(ipx/\hbar)
\label{plane}
\eeq
where the identification of the normalization factor $\langle 0|0\rangle$ as
$(2\pi\hbar)^{-1/2}$
stems from \eq{delta}. Using \eq{plane} it is easy to
show\cite{dirac,messiah} that
\beq
\langle x'|\hat{p}|x\rangle=-i\hbar{\partial\over\partial x}\delta(x-x')~,
\label{momentum}
\eeq
and that
\beq
[\hat{x},\hat{p}]=i\hbar.
\nonumber\eeq


\section{The time-energy transformation and the quantum evolution
equation}
\label{te}

We now wish to extend the proof presented in the previous section to the
time-energy domain. In classical mechanics
the use of canonical transformations to treat the time variable as a
coordinate is well established\cite{dirac1}. As discussed above,
the situation in quantum
mechanics is not so straightforward because one has to be careful
about the type of time that can be treated as an operator. We now
show that there are no problems associated with treating the
{\it ``proper time''}
as a coordinate and quantizing it.

For an inertial frame, in which the velocity is constant,
the classical ``proper-time'' is given\cite{goldstein1} as
\beq
t_s=t(1-v^2_s/c^2)^{1\over 2}~,
\label{cts}
\eeq
where $t$ is the time in
(an arbitrarily chosen) ``reference'' Lorentz frame
and $v_s\equiv dx_s/dt$ is the velocity of the $s$
frame {\it relative} to the reference frame.
Alternatively we could work with the Lorentz transformed time
\beq
t_s=(t-x_sv_s/c^2)/(1-v^2_s/c^2)^{1\over 2}~.
\label{cts1}
\eeq
We can define a self-adjoint bounded ``proper-time'' operator
by replacing $v_s$ in \eq{cts} by the velocity operator $\hat{v}_s.$
Thus,
\beq
\hat{t}_s=t(1-\hat{v}^2_s/c^2)^{1\over 2}~.
\label{ts}
\eeq

The arguments presented below are independent of the
exact form of
the velocity-squared $\hat{v}^2_s$ operator.
However for completeness we may,
following the classical relativistic
relation between the velocity and momentum,
\beq
v^2_s/c^2=p_s^2c^2/(p_s^2c^2+E_s^2),
\nonumber
\eeq
where $E_s=m_sc^2$ is the rest energy of body $s,$
define such an operator. In order to guarantee its self-adjointness,
we define a $\hat{v}_s^2$ operator
within the usual rigged Hilbert Space formulation of
quantum mechanics\cite{rigged,madrid}, as,
\beq
\hat{v}_s^2/c^2={1\over 2}\left[
\hat{p}_s^2c^2(\hat{p}_s^2c^2+\hat{E}_s^2)^{-1}+
(\hat{p}_s^2c^2+\hat{E}_s^2)^{-1}\hat{p}_s^2c^2\right].
\eeq

For fixed $t$
there are uncertainties associated with the
proper-times
due to the quantum mechanical uncertainties in $\hat{v}_s$.
We also note that for $t>0$, the spectrum of
$\hat{t}_s$ is bounded from above by $t$ (when $v_s=0$) and from
below by 0 (when $v_s=c$), with the upper bound becoming the
lower bound and {\it vice versa} when $t<0$.
The ``proper-time'' operator does not suffer from
singularities when the momentum $p_s=0$, besetting
such operators as the non-relativistic ``measurement-time'' or
``arrival-time''\cite{aharonov,paul,muga,grot}
or the ``tempus'' operator\cite{navarro}.

It follows from \eq{cts} that $\dot{t}_s$,
the classical {\it ``velocity of time''} is given as,
\beq
\dot{t}_s\equiv dt_s/dt=(1-\beta_s^2)^{1\over 2}.
\nonumber\eeq
We note that for a constant velocity the transformed time of \eq{cts1}
yields the same form for the ``velocity of time''.

The Lagrangian of \eq{rellag}
of two non-interacting $A$ and $B$ particles
can be written as
\beq
{\cal L}=-E_A\dot{t}_A-E_B\dot{t}_B~,
\nonumber\eeq
where $E_{A(B)}=m_{A(B)}c^2$ is the rest-energy of particle $A (B)$ in the
reference frame.
The classical (4$^{\rm th}$) momentum conjugate to $t_s$ is obtained by the
usual definition as
\beq
p_{t_s}=\partial{\cal L}/\partial \dot{t}_s=-E_s,~s=A,B~.
\nonumber\eeq

We can now canonically transform the proper-times
of bodies $A$ and $B$ to two new variables, defined as
$t_{1(2)}=(t_A\pm t_B)/2~.$ Naturally
$\dot{t}_{1(2)}=(\dot{t}_A\pm\dot{t}_B)/2~.$
Hence when we choose $E_A=E_B=E_s$
\beq
{\cal L}=
-E_s(\dot{t}_A+\dot{t}_B)=-2E_s\dot{t}_1
\nonumber\eeq
Therefore, the conjugate momenta to $t_1$ and $t_2$ are given as,
\beq
p_{t_1}=\partial{\cal L}/\partial \dot{t}_1=-2E_s~~~~~~~
p_{t_2}=\partial{\cal L}/\partial \dot{t}_2=0~.
\nonumber\eeq

The invariant we are now seeking is simply $-2E_s$
since
$$ -2E_s=p_{t_A}+p_{t_B}=p_{t_1}+p_{t_2}.
$$
Because $E_s$ is a function of
$p_{t_A}$ and $p_{t_B}$ and also of $p_{t_1}$ and $p_{t_2}$,
$\hat{E_s}$, the (rest-energy) operator corresponding to it,
commutes with all four momentum operators. Hence we can
write the eigenstates of
$-2\hat{E_s}$ in two different ways,
\beq
\langle t_A|p_{t_A},n\rangle\langle t_B|p_{t_B},n\rangle=
\langle t_1|p_{t_1},n\rangle
\langle t_2|p_{t_2},n\rangle~,
\label{1t}
\eeq
where $n$ is any other quantum number needed to specify the state. In the
present construction we choose on purpose just a single $n$ quantum number,
which therefore remains the same upon execution of the canonical
transformation. Henceforth we omit the explicit mention of $n$ but it
should be considered as present.

Choosing two particular eigenvalues, $p_{t_A}=p_{t_B}\equiv -E_s$
we have that
$p_{t_1}=-2E_s~,p_{t_2}=0~,$
with \eq{1t} now reading,
\beq
\bra{t_A}-E_s \rangle\bra{t_B}-E_s \rangle=\bra{(t_A+t_B)/2}-2E_s \rangle
\bra{t_A-t_B}0\rangle
~.
\nonumber
\eeq
If we now also choose the particular values $t_A=t_B\equiv t_s$ we have that
\beq
\bra{t_s}-E_s \rangle\bra{t_s}-E_s \rangle=
\bra{t_s}-2E_s \rangle\bra{0}0\rangle
~.
\nonumber
\eeq
we obtain that
\beq
\langle t_s|-2E_s \rangle'
=\Big(\langle t_s|-E_s \rangle'\Big)^2~,
\nonumber\eeq
where
\beq
\langle t_s|-E_s \rangle'
\equiv \langle t_s|-E_s \rangle/\langle 0|0\rangle.
\nonumber
\eeq
It immediately follows, by the same type of arguments presented in
the previous section that
\beq
\langle t_s|-E_s \rangle=a\exp(-E_s f(t_s)),
\label{ft}
\eeq
where $f(t_s)$ is only a function of $t_s$.

If instead of
choosing $E_A=E_B=E_s$, we now choose $E_A=E_s$ and $E_B=0$ (e.g., the
second particle is a photon)
we have that
\beq
{\cal L}=-E_s\dot{t}_A=-E_s(\dot{t}_1+\dot{t}_2),
\nonumber\eeq
and that
\beq
p_{t_B}=0,~{\rm and}~p_{t_A}=p_{t_1}=p_{t_2}=-E_s~.~~
\nonumber\eeq
It follows that
\beq
\langle t_A|-E_s \rangle\langle t_B|0\rangle=\langle t_1|-E_s \rangle
\langle t_2|-E_s \rangle~.
\nonumber
\eeq
Choosing $t_A=2t_s$ and $t_B=0$ we have that $t_1=t_s$ and $t_2=t_s$,
hence
\beq
\langle 2t_s|-E_s \rangle\langle 0|0\rangle=\langle t_s|-E_s \rangle
\langle t_s|-E_s \rangle=\Big(\langle t_s|-E_s \rangle\Big)^2~.
\nonumber
\eeq
We now obtain that
\beq
\langle t_s|-E_s\rangle=a\exp(t_s g(-E_s)).
\label{ge}
\eeq
By equating \eq{ft} and \eq{ge}
we obtain that $-g(-E_s)/E_s=f(t_s)/t_s=\alpha$, a constant,
which means that $g(-E_s)=-\alpha E_s$ and
\beq
\langle t_s|E_s \rangle=
a\exp(-\alpha E_s t_s).
\label{alpha}
\eeq

We now show that the constant $\alpha$ must be a purely imaginary number.
Because the $[0,t]$ boundaries of $t_s$ are {\it finite},
we cannot argue, as we did in the coordinate-momentum case of
the previous section,
where the boundaries were infinite,
that $\alpha$ must be a purely imaginary number
because otherwise the amplitude
would diverge at the boundaries.
Rather, in the present case, $\alpha$ must
be purely imaginary to maintain Lorentz invariance. This is because
if $\alpha$ had a real part
the $|\langle t_s|E_s \rangle|=|a|\exp(-R_e(\alpha) E_s t_s)$
distribution would be maximal (minimal) at $t_s=t,$ for
$R_e(\alpha)>0~ (R_e(\alpha)<0),$
giving rise
to different physical observations for different definitions of $t$, i.e.,
different reference frames, in clear contradiction to Lorentz invariance.
Thus the constant $\alpha$ must be a purely imaginary number
$\alpha=i\gamma$.
By again
identifying $\gamma$ with $1/\hbar$, (which is in fact simply the
definition of the scaling of $E_s$) we obtain that,
\beq
\langle t_s|E_s \rangle=
\langle 0|0\rangle e^{-iE_st_s/\hbar}.
\label{ept}\eeq

\noindent
Three comments are now in order:
\newline
1)
The above derivation applies even to a truly
structureless elementary particle for which the rest energy is just a
single number. The reason is that in this case \eq{ge}
still holds
because $t_s$ is definitely a continuous variable and the infinitesimal
change in it, which is part of our proof, is perfectly permissible.
Because there is now only a single value of $E_s$, the function $g(-E_s)$
now becomes a simple number $g_s,$ and we define $E_s$ as
$E_s=-g_s\hbar,$ with \eq{ept} immediately following.

\noindent
2)
The above proof also holds for bound states.
In this case the energies may be varied
by subjecting the particle to an external field and changing the strength
of this external field. Once \eq{ept} is proved in the
presence of the field, we can adiabatically switch off the field while
establishing \eq{ept} for smaller and smaller external fields, until
we reach the limit when the external field is zero.

\noindent
3)
The spectrum of $\hat{t}_s$, depending on the spectrum of $\hat{v}_s$,
extends from $t_s=0$ to $t_s=t$. As pointed out by Galapon\cite{galapon},
for such a bounded time operator, in the rigged Hilbert Space
of scattering theory\cite{rigged,madrid} the objections raised by
Pauli\cite{pauli} against the existence of a self-adjoint time operator
(having to do with the non-existence of a conjugate Hamiltonian operator
with a point spectrum) do not apply.

We can now use \eq{ept},
to construct the rest-energy operator
in the proper-time representation as,
$$
\langle t_s|\hat{E}_s|\Psi\rangle=
$$
$$\int dt'\langle t_s| \left\{\sum_i E_i|E_{i}\rangle\langle E_{i}|
+\int dEE|E\rangle\langle E|\right\}|t'\rangle\langle t'|\Psi\rangle=
$$
$$
|a|^2\int dt'\left\{\sum_i E_i
e^{-iE_{i}(t_s-t')/\hbar}
+\int dEEe^{-iE(t_s-t')/\hbar}\right\}
\langle t'|\Psi\rangle
$$
$$
=i\hbar{\partial\over\partial t_s}
\int dt'\langle t_s|
\left\{\sum_i
|E_{i}\rangle\langle E_{i}|
+\int dE|E \rangle\langle E|\right\}|t'\rangle
\langle t'|\Psi\rangle
$$
\beq
=i\hbar{\partial\over\partial t_s}
\int dt'\langle t_s|t'\rangle
\langle t'|\Psi\rangle=
i\hbar{\partial\over\partial t_s}\langle t_s|\Psi\rangle~,
\label{evol}
\eeq
where $E_i$ are the discrete eigenvalues of $\hat{E}_s$.
The $[\hat{t}_s,\hat{E}_s]=-i\hbar$ commutation relation
follows immediately from \eq{evol},
leading in the usual fashion\cite{messiah} to the corresponding
uncertainty relations.

The $t_s=t$ eigenvalue is of special
interest because it occurs when $v_s=0$, i.e., when we
equate the $s$-inertial frame with the reference frame. In this case
$\hat{E}$ - the total-energy operator
and $\hat{E}_s$ the rest-energy operator coincide.
We thus have as a special case of \eq{ept} that
\beq
\langle t|E\rangle=ae^{-iEt/\hbar}.
\label{fet}
\eeq
and as a special case of \eq{evol}
that
\beq
\langle t|\hat{E}|\Psi\rangle=
i\hbar{\partial\over\partial t}\langle t|\Psi\rangle~.
\label{evolt}
\eeq
Note however that $t$ in contrast to $\hat{t}_s$ is a number and not
an operator.
We have thus obtained the quantum mechanical evolution equations
(\eq{energy}).

In special relativity,
the substitution of the
$\hat{E}^2=\hat{E}^2_s+\hat{p}_s^2c^2$ on the r.h.s. of \eq{energy},
together with \eq{momentum}, leads to the time dependent
Klein-Gordon\cite{sakurai} or Dirac\cite{dirac,sakurai} equations.
In the non-relativistic limit the substitution
of $\hat{E}=\hat{p}^2/2m$ together with
\eq{momentum} leads to the time-dependent Schr\"odinger equation.

\vskip .2truein
\section{Extensions to accelerating systems}

So far we have treated inertial systems in which {$v_s$} was constant.
For accelerating systems for which {$v_s(t)$} is a non-constant function of time,
the definition of the proper time must be modified. Realizing that the
speed of light in any frame must still be conserved, the differential relation
\beq
dt_s=dt(1-v_s^2(t)/c^2)^{1\over 2}
\label{dts}
\eeq
still holds.
This means that
{\beq
t_s=\int_0^tdt'(1-v_s^2(t')/c^2)^{1\over 2}.
\eeq}
Alternatively we can consider the differential transformed time
\beq
dt_s=(dt-dx_sv_s(t)/c^2)/(1-v_s^2(t)/c^2)^{1\over 2}
\eeq
In either case, the {\it velocity of time}
{$\dot{t}_s$} remains the same because in either differential equation for
{$dt_s$}, we have that
\beq\dot{t}_s=(1-v_s^2(t)/c^2)^{1\over 2}.\eeq
Thus our entire analysis expressed in proper time and rest
energy is correct for
non-inertial systems as is and we obtain that
\beq
\langle t_s|\hat{E}_s|\Psi\rangle=
i\hbar{\partial\over\partial t_s}\langle t_s|\Psi\rangle~.
\label{generalr}
\eeq
However we can no longer make the transition to the total-energy and the
``time" because the system linked to the particle is no longer an inertial
system and cannot serve as a ``reference frame".

We note that for accelerating frames,
$\langle t_s|E_s\rangle$, although remaining the same as a function of
$t_s$, assumes a different form as a function of $t$. We have that
\pagebreak
$$
\langle t_s|E_s\rangle=
a\exp(-iE_st_s/\hbar)=
$$
$$
a\exp\left(-iE_s\int_0^tdt'(1-v_s^2(t')/c^2)^{1\over 2}/\hbar\right)=
$$
\beq
a\exp\left(-i\int_0^tdt'{\cal L}(t')/\hbar\right).
\eeq
The motion no longer represents a plane wave in space-time but a motion
in a curved space!
It is instructive to derive a semi-classical relativistic
equation for the motion of an accelerating particle. Using the relation
$E(t')=E_s\left[1-v_s^2(t')/c^2\right]^{-{1\over 2}}$ we have that
$$
\exp(-iE_st_s/\hbar)=
\exp\left(-i\int_0^tdt'E(t')\left[1-v_s^2(t')/c^2\right]/\hbar\right)=
$$
\beq
\exp\left(-i\int_0^tdt'E(t')/\hbar\right)\exp\left(i\int_0^tdt'
E(t')v_s^2(t')/(c^2\hbar)\right).
\eeq
Dropping the $s$ index as applying to the velocity and momentum, and
using the definition $v(t')=dx'/dt'$ and the classical
relativistic relation
$p(x')=E(t')v(t')/c^2,$ we have that
$$
\exp(-iE_st_s/\hbar)=
$$
\beq
\exp\left(-i\int_0^tdt'E(t')/\hbar\right)
\exp\left(i\int_0^x dx'p(x')/\hbar\right).
\label{spatem}
\eeq
We have thus derived a joint spatio-temporal semiclassical form of the
relativistic wave function in curved spaces.

\section{Discussion}
\label{discussion}
In this paper we have derived the special-relativistic
energy-time transformation
matrices and the quantum evolution equations. A dynamical quantum mechanical
equation valid for accelerating frames was also derived.
The derivation involved
the {\it proper-time} operator, defined
as a function of $\hat{v}_s$, the velocity operator, and depending
parametrically on $t$, the time measured in a selected inertial frame,
the so-called ``reference-frame''.

The present derivation is the only one to our
knowledge where the quantum dynamical equations are derived from
{\it within} conventional quantum mechanics, making no new postulates.
It relies only on the basic structure of the rigged Hilbert space of
quantum mechanics;
on the linear correspondence between observables and quantum operators;
on canonical invariance with respect to transformation between
pairs of conjugate variables; and on Lorentz invariance.
It is interesting to note that the only aspect of canonical invariance
used here is that the canonically transformed conjugate pair of variables
describe the same point in
phase space as the untransformed conjugate pair.

Applications of Eqs. (\ref{generalr}-\ref{spatem})
to quantum dynamics
of highly accelerating systems
subject to gravitational potentials
are considered elsewhere\cite{generalr}.

\vskip .1truein

\centerline {\bf Acknowledgments}

\vskip .1truein

Various discussions with P. Brumer, and the assistance of E.A. Shapiro, and
I. Thanopulos are gratefully
acknowledged.


\end{document}